\documentclass[a4paper,fleqn,usenatbib]{mnras}

\usepackage{newtxtext,newtxmath}

\usepackage[T1]{fontenc}
\usepackage{ae,aecompl}

\usepackage{graphicx}	
\usepackage{amsmath}	
\usepackage{amssymb}	
\DeclareMathOperator{\sech}{sech}

\title[General model for vertical disc distribution]{General model of vertical distribution of stars in the Milky Way using complete Jeans equations}

\author[S. Sarkar \& C.J.Jog]{
Suchira Sarkar,$^{1}$\thanks{Contact e-mail: \href{suchira:suchira@iisc.ac.in}{suchira@iisc.ac.in}}
Chanda J.Jog$^{1}$
\\
$^{1}$Department of Physics, Indian Institute of Science, Bangalore-560012, India
}

\date{Accepted XXX. Received YYY; in original form ZZZ}

\pubyear{2019}

\begin{document}
\label{firstpage}
\pagerange{\pageref{firstpage}--\pageref{lastpage}}
\maketitle

\begin{abstract}
The self-consistent vertical density distribution in a thin, isothermal disc is
typically given by a $\sech^{2}$ law, as shown in the classic work by Spitzer (1942). This is obtained assuming that the radial and vertical
motions are decoupled and only the vertical term is used in the Poisson equation. We argue
that in the region of low density as in the outer disc this treatment is no longer
valid. We develop a general, complete model that includes both radial and vertical
terms in the Poisson equation and write these in terms of the full radial and vertical
Jeans equations which take account of the non-flat observed rotation curve, the random motions,
and the cross term that indicates the tilted stellar velocity ellipsoid. We apply it to the Milky Way and show that 
these additional effects change the resulting density distribution significantly,
such that the mid-plane density is higher and the disc thickness (HWHM) is lower by 30-40\% in the outer Galaxy. Further, the vertical
distribution is no longer given as a $\sech^{2}$ function even for an isothermal case. 
These predicted differences are now within the verification limit of  new, high-resolution  data for
example from GAIA and hence could be confirmed.  
\end{abstract}


\section{Introduction}

The vertical structure of the stellar disc in a thin galactic disc such as the Milky Way is typically obtained by treating it as an isothermal, self-gravitating system 
whose density  distribution along the vertical  direction
is 
given by a $\sech^{2}$ form, see the classic work by Spitzer (1942). The disc is treated to be  thin and axisymmetric; and a cylindrical coordinate system is used. Hence the Poisson equation contains only the vertical term, while the radial and azimuthal terms are dropped. Combining it with the vertical Jeans equation 
or the equation of hydrostatic equilibrium, gives 
the following equation
\begin{equation}
\frac{\partial}{\partial z}\left[\frac{1}{\rho}\frac{\partial(\rho \overline{v^{2}_{z}})}{\partial z}\right] = -4 \pi G \rho,\label{eq:1}
\end{equation}

\noindent where $\rho$ is the density and $\overline{v^{2}_{z}}$ is the square of the isothermal vertical velocity dispersion. 
The solution of this equation gives the 
vertical density distribution:
\begin{equation}
\rho(z) = \rho_{0}\sech^{2}(z/z_{0})\label{eq:2}
\end{equation}

\noindent where $\rho_{0}$ is the mid-plane density and $z_{0}$ is a measure of scaleheight.  We will refer to this as {\it the $\sech^{2}$ model} in rest of the paper. 
A similar treatment is used to study the related problem of the total dynamical mid-plane density or the Oort limit (Oort 1932; Bahcall 1984). 
Thus, usually in galactic dynamics, the $R$ and $z$ motions are effectively taken to be decoupled. This allows for a simple treatment of the orbits in the plane and 
those normal to the plane as being decoupled (Mihalas \& Routly 1968; Binney \& Tremaine 1987).

We point out that in the general case, such as when the disc density is low as in the outer parts of the disc or at regions away from the mid-plane, or as in the thick disc, the neglect of the radial term 
is not justified because in such regions this term may become comparable to the vertical term in the Poisson equation. Hence even for an isothermal case, the resulting density distribution would be different from the standard $\sech^{2}$ distribution.

The radial term in the Poisson equation was indeed included, and for simplicity was obtained using the observed rotation curve, in some earlier papers (Narayan, Saha \& Jog, 2005; Banerjee et al.,2011; Sarkar \& Jog, 2019). Interestingly, in case of an observed flat rotation curve, the contribution of this term turns out to be identically zero (Narayan, Saha \& Jog, 2005) in the mid-plane. Here instead we use the complete radial Jeans equation to calculate the radial term in the Poisson equation.  

Further, in writing both the vertical and radial Jeans equations that are used to obtain the vertical and radial terms of the Poisson equation  
respectively, typically a simplification is made, namely the so-called cross term $\overline{v_{R}v_{z}}$, where the average is taken over the velocity dispersions, is ignored. The term $\overline{v_{R}v_{z}}$ is a component of the velocity ellipsoid tensor whose value is set by the tilt of the velocity ellipsoid w.r.t. 
the disc plane, and it also represents the coupling between radial and vertical motion. But its contribution has been neglected in most of the dynamical models so far (Bahcall 1984; Cappellari 2008) as is justified for a region very close to the mid-plane in a thin disc. Note that recent kinematic data from observations like RAVE, SDSS, and  GAIA show that the stellar velocity ellipsoid is indeed tilted in the meridional plane both in the stellar disc (Hagen et al.,2019; Evarall et al.,2019) and in the outer stellar halo (Wegg et al.,2019). It is also observed to have an orientation such that it tends to align with a spherical polar coordinate system centered at the center of the Galaxy. This motivates us to consider this cross term in the Jeans equations. 

Here, we attempt to solve for the vertical distribution of stars using the complete Poisson equation, where the vertical and radial terms are given in terms of  complete vertical and radial Jeans equations without the usual simplifications - thus we can consider this as the complete and general model. We calculate the mid-plane density value and scaleheight of the disc, measured in terms of the half-width at half-maximum of the vertical density distribution, and compare them with the corresponding results obtained from a simple, stars-alone case ($\sech^{2}$ model) as obtained from our earlier work (Sarkar \& Jog, 2018). With the advent of accurate data, as from GAIA or LAMOST, it is feasible to check such modified density distribution with the observations, hence our study is timely.
 Some earlier studies have used the full set of Jeans equations to determine vertical force field, shape of the dark matter halo, and local estimate of dark matter density (Bovy \& Tremaine, 2012; Hagen \& Helmi, 2018; S\'{a}nchez-Salcedo et al.,2016; Wegg et al.,2019), but the effect on the vertical density distribution of stars has not been studied. 

The outline of the paper is as follows- we describe the derivation of the equations along with the input parameters in section 2, give the results in section 3, and give the conclusions in section 4. 

\section{Formulation of the problem}
\subsection{Derivation of the equations}

Here we use the full Poisson equation containing both the vertical and radial terms, where the vertical and radial gradients of potential are obtained using the full Jeans equations. We consider the disc to be in a steady-state system and for such a system the Poisson equation is given by

\begin{equation}
\frac{1}{R}\frac{\partial}{\partial R}\left(R \frac{\partial \Phi}{\partial R}\right)+\frac{\partial^{2}\Phi}{\partial z^{2}} = 4 \pi G \rho(R,z)\label{eq:3}
\end{equation} 

\noindent where $\Phi$ is the gravitational potential, and the terms on the L.H.S denote the radial and vertical terms of the Poisson equation respectively. To calculate the vertical and radial gradients of the potential, we use the radial and vertical axisymmetric Jeans equations in cylindrical coordinates, given as (see Binney \& Tremaine, 1987):

\begin{equation}
\begin{split}
\frac{\partial \Phi}{\partial R} &= -\frac{1}{\rho}\frac{\partial}{\partial R}(\rho \overline{v^{2}_{R}})-\frac{1}{\rho}\frac{\partial}{\partial z}(\rho \overline{v_{R}v_{z}})-\frac{(\overline{v^{2}_{R}}-\overline{v^{2}_{\phi}})}{R},\label{eq:4}
\end{split}
\end{equation}

\begin{equation}
\frac{\partial \Phi}{\partial z} = -\frac{1}{\rho R}\frac{\partial}{\partial R}(R \rho \overline{v_{R}v_{z}})-\frac{1}{\rho}\frac{\partial}{\partial z}(\rho \overline{v^{2}_{z}}).\label{eq:5}
\end{equation}

\noindent We consider that there is no net streaming motion along radial and vertical directions in the Galaxy, hence we can write $\overline{v^{2}_{R}}=\sigma^{2}_{R}$ and $\overline{v^{2}_{z}}=\sigma^{2}_{z}$ where $\sigma_{R}$, $\sigma_{z}$ represent velocity dispersions of stars along $R$ and $z$ directions respectively. $\overline{v_{\phi}^{2}}$ can be written as $\overline{v_{\phi}^{2}}=\sigma^{2}_{\phi}+\overline{v_{\phi}}^{2}$, where $\overline{v_{\phi}}$ represents observed mean rotation of stars in the disc and $\sigma_{\phi}$ is the azimuthal velocity dispersion.

At z=0, the mid-plane of the Galaxy, the velocity ellipsoid remains perfectly aligned with the cylindrical coordinate axes, which makes $\overline{v_{R}v_{z}} =0$ (Mihalas \& Routly, 1968). But away from the mid-plane, i.e, for $z\neq 0$ the ellipsoid can be shown to be tilted w.r.t the coordinates axes (Mihalas \& Routly,1968) and such a deviation gives rise to a non-zero $\overline{v_{R}v_{z}}$ term, as also found in observations (Sec.1). 
Recently, such a tilted ellipsoid was included in the studies by Cappellari (2019), and Nitschei et al. (2019) who solved the Jeans equations to obtain the kinematical quantities and compared these with observations. These papers considered the tilted velocity ellipsoid which was taken to be aligned along the spherical system for $z\neq 0$. However since they solved the equations in spherical co-ordinates the cross term drops out of the calculation. We note that these papers did not solve the joint Poisson equation and the Jeans equations to obtain the vertical density distribution of stars as done in our work.

Here we write the Jeans equations in cylindrical coordinates and consider the velocity ellipsoid to be aligned with spherical coordinates, centered at the center of the Galaxy, at all radii for $z\neq 0$. Thus we include the effect of the cross-term. We use the expression for $\overline{v_{R}v_{z}}$ given as $(\sigma_{R}^{2}-\sigma_{z}^{2})(z/R)$ (Mihalas \& Routly, 1968; Binney \& Tremaine, 1987). Using eqs. 4 \& 5 we calculate the radial and vertical terms of the Poisson equation to be:

\begin{equation}
\begin{split}
\frac{1}{R}\frac{\partial}{\partial R}\left(R\frac{\partial \Phi}{\partial R}\right) &= - \frac{1}{R}\frac{\partial}{\partial R}\left[\frac{R}{\rho}\frac{\partial}{\partial R}(\rho\sigma_{R}^{2})\right] \\
									  & - \frac{1}{R}\frac{\partial}{\partial R}\left[\frac{1}{\rho}\frac{\partial}{\partial z}({\rho z(\sigma_{R}^{2}-\sigma_{z}^{2}))}\right] \\
									  &  -\frac{1}{R}\frac{\partial \sigma_{R}^{2}}{\partial R}+\frac{1}{R}\frac{\partial}{\partial R} (\sigma^{2}_{\phi}+\overline{v_{\phi}}^{2}) 
\end{split}\label{eq:6}
\end{equation}

\begin{equation}
\begin{split}
\frac{\partial^{2} \Phi}{\partial z^{2}} &= -\frac{\partial}{\partial z}\left[\frac{1}{\rho R}\frac{\partial}{\partial R}(\rho z(\sigma_{R}^{2}-\sigma_{z}^{2}))\right]-\frac{\partial}{\partial z}\left[\frac{1}{\rho}\frac{\partial}{\partial z}(\rho \sigma_{z}^{2})\right]\label{eq:7}
\end{split}
\end{equation}

\noindent We consider an exponential stellar disc whose surface density falls off as $\Sigma=\Sigma_{0}\exp(-R/R_{D})$. Similarly, we also consider the radial velocity dispersion to fall off exponentially as $\sigma_{R}=\sigma_{0}\exp(-R/R_{vel})$ and define the vertical and azimuthal dispersions in terms of ratio to $\sigma_{R}$, e.g., $b_{z}=\sigma^{2}_{z}/\sigma^{2}_{R}$ and $b_{\phi}=\sigma^{2}_{\phi}/\sigma^{2}_{R}$.

Using  the above physical assumptions, and expanding eqs.(6,7) and substituting into eq.(3) we obtain the second-order differential equation describing the vertical density distribution as:

\begin{equation}
\begin{split}
\frac{d^{2}\rho}{dz^{2}}  & = \frac{-4 \pi G \rho^{2}}{\sigma_{z}^{2}}+\frac{1}{\rho}\left(\frac{d\rho}{dz}\right)^{2}+\frac{\rho}{\sigma_{z}^{2}}\frac{2\overline{v_{\phi}}}{R}\left(\frac{\partial\overline{v_{\phi}}}{\partial R}\right)\\
                          &   -\frac{z(1-b_{z})}{b_{z}R}\left(\frac{2}{R_{D}}-\frac{2}{R_{vel}}\right)\frac{d\rho}{dz} \\		
                          &   +\frac{\rho}{b_{z}}\bigg[\frac{2}{RR_{vel}}-\frac{4}{({R_{vel}})^{2}}+\frac{1}{RR_{D}}-\frac{2}{R_{D}R_{vel}}+\frac{4(1-{b_{z}})}{RR_{vel}} \\
                          &   \qquad +\frac{2(1-b_{\phi})}{RR_{vel}}+\frac{(1-{b_{z}})}{RR_{D}}\bigg] \label{eq:8}
\end{split}
\end{equation}
\noindent The first three terms of the above equation correspond to  eq.(1) and hence the rest of the terms in eq.(8) arise due to the various generalizations considered here. It allows us to study the effect of any kinematical term in a methodical way. The algebraic simplifications leading to the above equation consider $(1/\rho)(\partial \rho/\partial R)=-1/R_{D}$ which assumes that $\rho=\Sigma/2h_{z}$, where $h_{z}$ is the disc thickness at any radius. Though the disc thickness does not remain constant with radii (as shown later in sec.3.2), we expect that this approximation is justified as it will give rise to only second order effects in the results at a particular radius.

Equation 8 represents a general formalism and therefore can be applied for any typical disc galaxy satisfying the above physical assumptions. Here we have applied it for our Galaxy and we describe all the required input parameters in the following subsection.
\subsection{Input parameters and numerical solution}
For the surface density of the stellar disc of the Milky Way we have used $\mathrm{R_{D}=3.2 kpc}$, $\Sigma_{0}=640.9 M_{\odot}pc^{-2}$ 
(Mera et al.,1998). For $\sigma_{R}$ we consider $\mathrm{R_{vel}=8.7 kpc}$, $\sigma_{0}=105.0 km~s^{-1}$ 
as observed by Lewis \& Freeman (1989). We take $b_{z}$ to be $(0.45)^{2}$, consistent with the observation in the solar neighborhood (Dehnen \& Binney 1998; Mignard 2000) and assume this to be valid at all radii. 
These same input parameter values were used for the $\sech^{2}$ model (Sec.1). The azimuthal dispersion is related to radial dispersion as in epicyclic approximation, hence the ratio $b_{\phi}$ is calculated to be:
\begin{equation}
b_{\phi} = \frac{\sigma^{2}_{\phi}}{\sigma^{2}_{R}}  = \frac{1}{2}\frac{\left(V_{c}/R+dV_{c}/dR\right)}{V_{c}/R}\label{eq:9}
\end{equation} 
\noindent where $V_{c}$ is the circular velocity. We note that this expression is derived considering a circular orbit for stars and hence a true circular velocity curve should be used. But here we consider the mean rotation velocity $\overline{v_{\phi}}$ from the observed rotation curve, which deviates from the true circular velocity, to calculate the ratio. We expect that this approximation is justified, as it may lead to only second-order effect in the $b_{\phi}$ values. We use the observed rotation velocity data from Sofue (2012) and fit a polynomial to the observed data. Using the best-fit curve we calculate both the rotation velocity and its gradient at any radius and determine the ratio $b_{\phi}$ using eq. (9). 
The values obtained are 0.46, 0.54, 0.57, 0.56, 0.49, 0.41 at R=2, 3, 4, 5, 6, \& 7 kpc respectively. 
Starting from and beyond the solar neighborhood, the observed data shows large fluctuations, however, the theoretically fitted smooth curve in Sofue (2012) is nearly flat. Therefore, for $R \geq 8.5$ kpc, we consider the rotation curve to be flat for simplicity which gives $b_{\phi}$ identically = 0.5 from eq.(9).

We note that, as the velocity dispersion of stars falls with radius, its value may get less than the gas dispersion, beyond a certain radius. Since stars are formed from gas clouds, hence they cannot have a lower velocity dispersion than the gas itself, as explained in Sarkar \& Jog (2018). The HI velocity dispersion in the outer Galaxy is observed to saturate around 7 kms$^{-1}$ (Kamphuis 1993; Dickey 1996). Therefore, we keep the stellar vertical and azimuthal dispersion values constant at 7.5 km s$^{-1}$ from R=18 kpc and 20 kpc onward respectively to keep it higher than the gas dispersion and update the values of $b_{\phi}$ and $b_{z}$ accordingly at those radii.  

\medskip
\noindent
We solve eq. (8) using the method as in Narayan \& Jog (2002), and Sarkar \& Jog (2019). That is, we solve it by applying the fourth-order Runge-Kutta method, where the observed surface density of stars is used as one boundary condition and the second condition is given by $d\rho/dz=0$, defined at z=0. This condition is satisfied for any realistic vertical density distribution which will be homogeneous close to the mid-plane. 

\section{Results: Vertical stellar disc structure}
\subsection{Effect on the mid-plane density $\rho_{0}$ of the stellar disc}

We solved for the vertical density distribution for two cases. First, when the disc structure is solved by writing
 the radial term in equation (3) in terms of the observed, non-flat rotation curve 
 from R=2 to 7 kpc, to study its effect distinctly. This is done by solving the following equation (Model A), as obtained in Sarkar \& Jog (2019):

\begin{equation}
\begin{split} 
\frac{d^{2}\rho}{dz^{2}}  & = \frac{-4 \pi G \rho^{2}}{\sigma_{z}^{2}}+\frac{1}{\rho}\left(\frac{d\rho}{dz}\right)^{2}+\frac{\rho}{\sigma_{z}^{2}}\frac{2\overline{v_{\phi}}}{R}\left(\frac{\partial\overline{v_{\phi}}}{\partial R}\right)
\end{split}\label{eq:11} 
\end{equation}
\noindent The observed mean rotation velocity values (Sofue, 2012) are used in the last term in the above equation.

\begin{table}
\caption{Mid-plane density $\rho_{0}$ using the $\sech^{2}$ model, model with only a non-flat, observed rotation curve (Model A) and the complete, general model (Model B)}
\label{table:1}
\centering
\begin{tabular}{l l l l l l}
\hline \hline
Radius & $(\rho_{0})_{\sech^{2}}$ & $(\rho_{0})_{\mathrm{model~A}}$ & $\Delta_{\mathrm{A}}$ & $(\rho_{0})_{\mathrm{model~B}}$ & $\Delta_{B}$  \\
(kpc) & $(M_{\sun}\mathrm{pc^{-3}})$ & $(M_{\sun}\mathrm{pc^{-3}})$ & \% & $(M_{\sun}\mathrm{pc^{-3}})$ & \% \\
\hline
2.0	&	0.568				&		0.604		&	+6.3	&	0.463				&	 -18.5 \\
3.0	&	0.381				&		0.365		&	-4.2	&	0.275				&	 -27.8 \\
4.0	&	0.258				&		0.236		&	-8.5	&	0.190				&	 -26.4 \\
5.0	&	0.173				&		0.162		&	-6.4	&	0.140				&	 -19.1\\
6.0	&	0.117				&		0.118		&	+0.85	&	0.108				&	 -7.7 \\
7.0	&	0.078				&		0.086		&	+10.2	&	0.081				&	 +3.8 \\
8.5	&	0.043				&		--		&	0.0	&	0.041				&	 -4.6 \\
10.0	&	0.024				&		--		&	0.0	&	0.023				&	 -4.2 \\
12.0	&	0.011020			&		--		&	0.0	&	0.011045			&	 +0.2 \\
14.0	&	$\mathrm{5.13\times 10^{-3}}$	&		--		&	0.0	&	$\mathrm{5.27\times 10^{-3}}$	&	 +2.7 \\
16.0	&	$\mathrm{2.38\times 10^{-3}}$	&		--		&	0.0	&	$\mathrm{2.55\times 10^{-3}}$	&	 +7.1 \\
18.0	&	$\mathrm{6.54\times 10^{-4}}$	&		--		&	0.0	&	$\mathrm{8.12\times 10^{-4}}$	&	 +24.2 \\
20.0	&	$\mathrm{2.17\times 10^{-4}}$	&		--		&	0.0	&	$\mathrm{3.03\times 10^{-4}}$	&	 +39.6	\\
22.0	&	$\mathrm{8.78\times 10^{-5}}$	&		--		&	0.0	&	$\mathrm{1.24\times 10^{-4}}$	&	 +41.2	\\
\hline	
\end{tabular}
\end{table}

\begin{figure*}
\centering
\includegraphics[height=2.3in,width=3.2in]{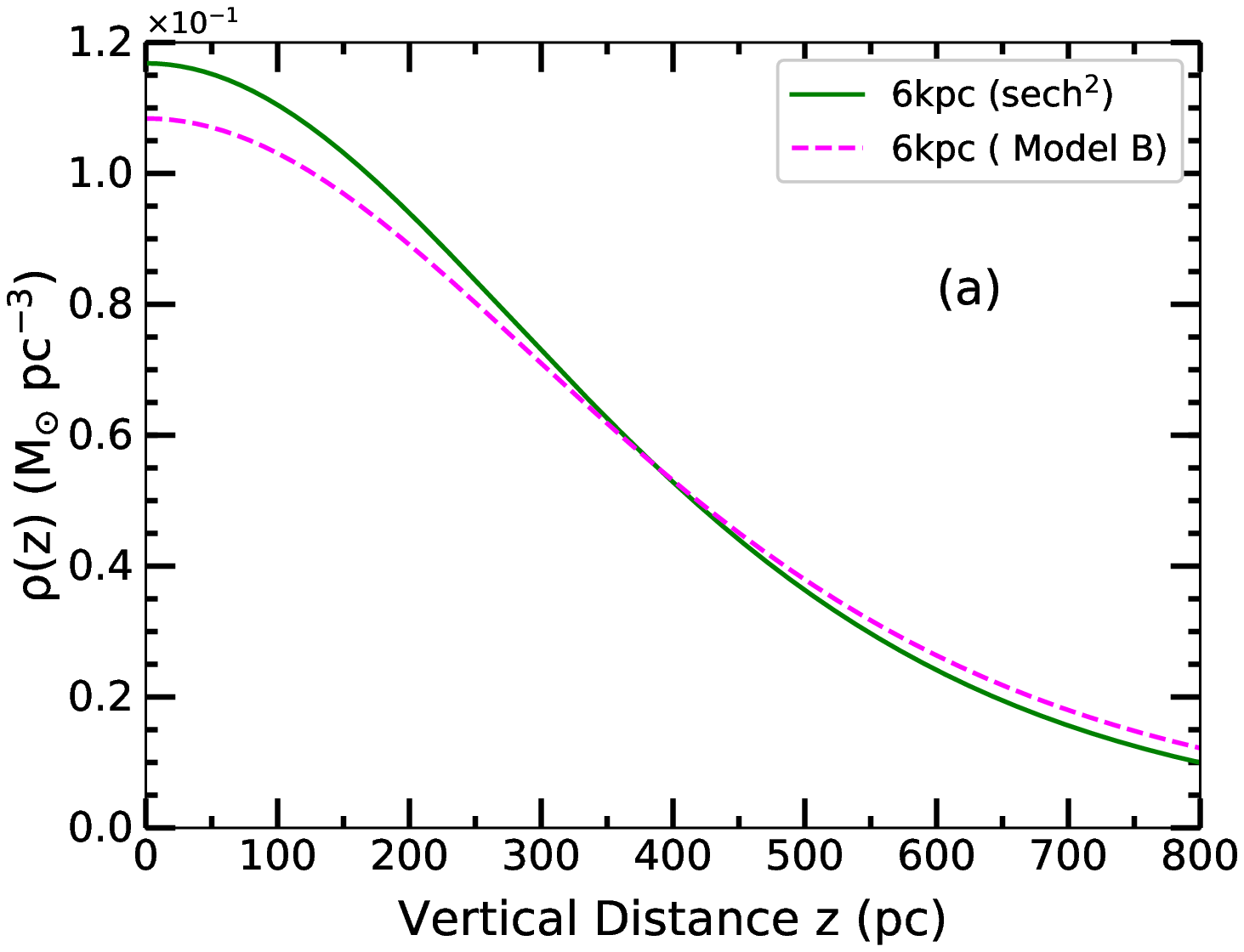}
\medskip
\includegraphics[height=2.3in,width=3.2in]{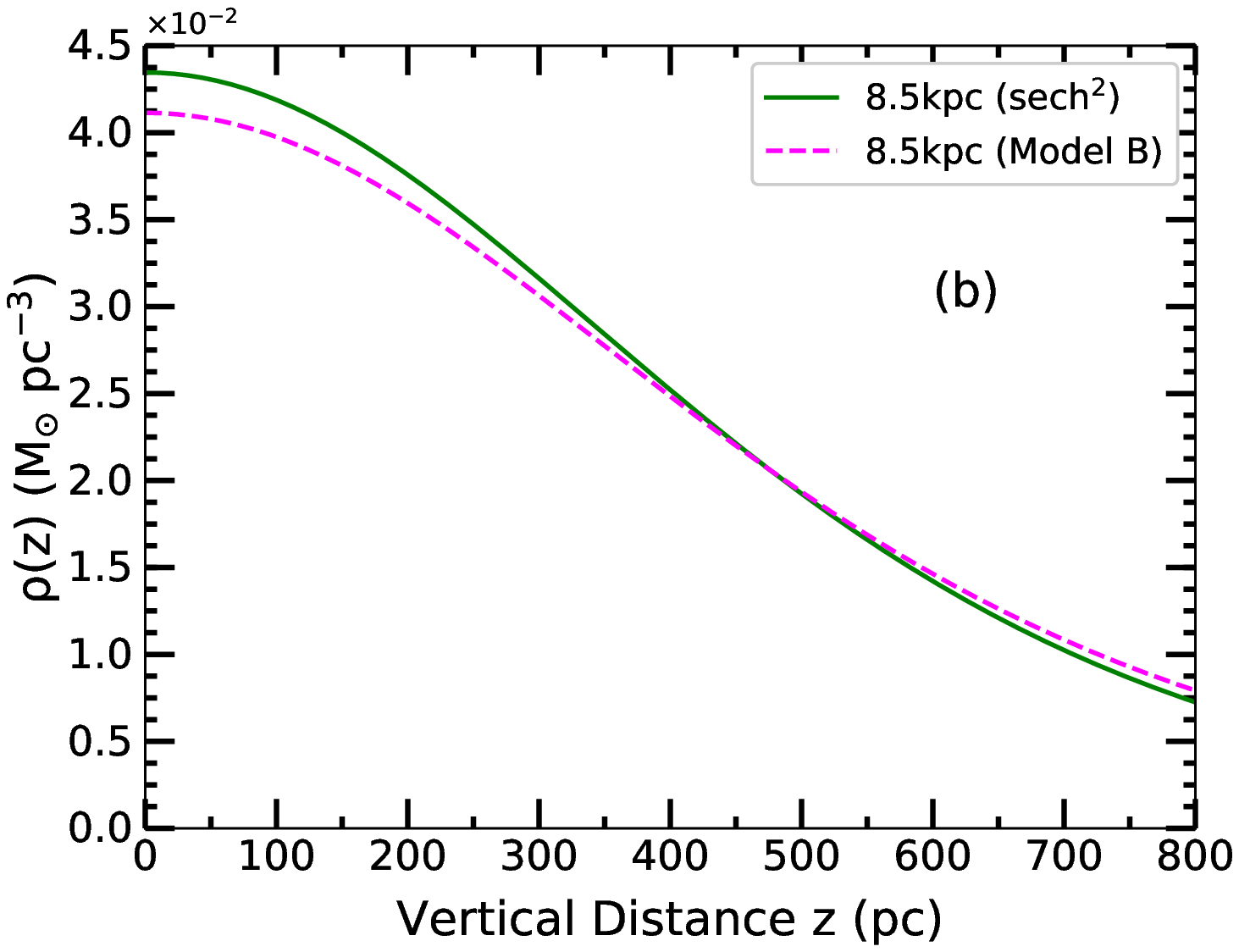}
\medskip
\includegraphics[height=2.3in,width=3.2in]{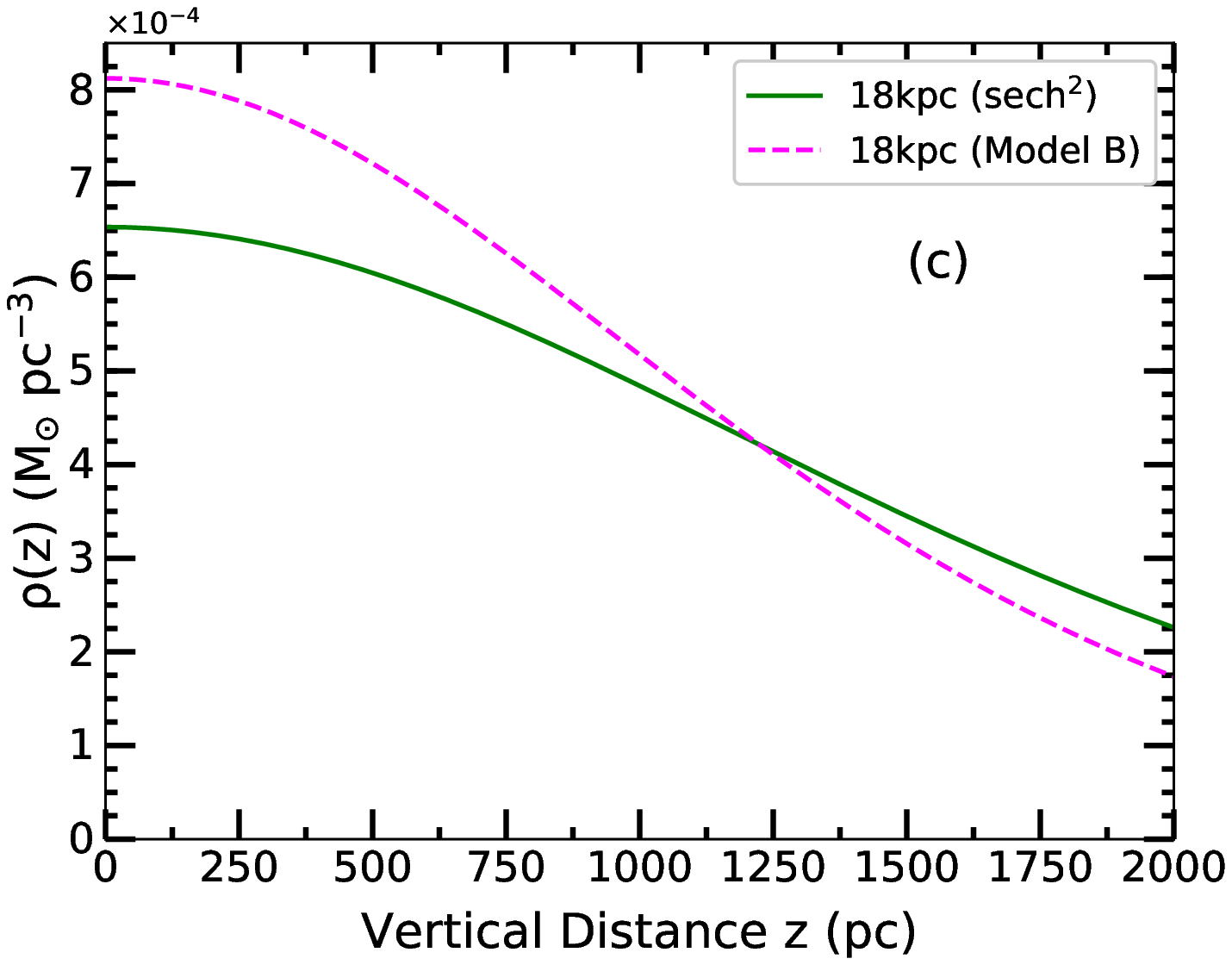}
\medskip
\includegraphics[height=2.3in,width=3.2in]{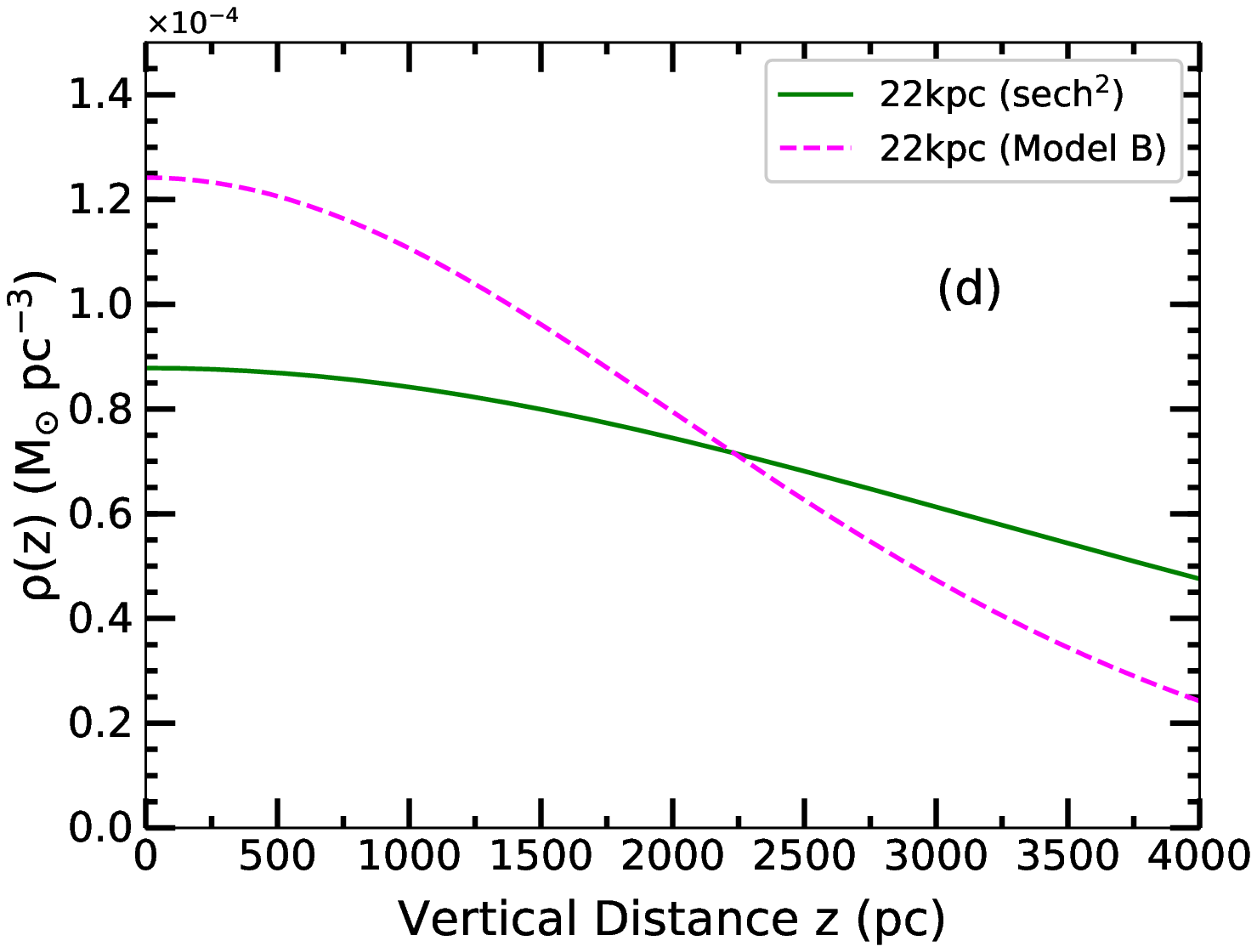}
\caption{Plot of vertical density distribution ($\rho(z)$) of stars at R=6, 8.5, 18 \& 22 kpc (a,b,c,d) respectively. The Green solid curves represent $\rho(z)$ for a $\sech^{2}$ distribution (see eq.(2) in sec.1) 
and the Magenta dashed curves represent $\rho(z)$ obtained using the complete, general model (model B) here. The difference between these two distributions is more prominent in the outer Galaxy and turns out to be very small at the 6 kpc and solar radius. This shows the significance of using the complete Poisson and Jeans equations while solving for $\rho(z)$ for stars, specially in the Outer Galaxy.}
\label{label1}
\end{figure*}

The resulting mid-plane density $\rho_{0}$ values are listed in Table 1, under "Model A" and are compared with the results from $\sech^{2}$ model (stars-alone case) obtained in Sarkar \& Jog (2018), with $\Delta_{A}$ giving the \% difference between these two models.
 We note that $\rho_{0}$ has increased at some radii (R=2,6,7 kpc) and has decreased at some radii (R=3,4,5 kpc) which happens due to the negative and positive gradient, respectively, present in the non-flat rotation curve at the corresponding radii. The last term in eq.(10), shows that the amount of change introduced in the ($\rho_{0}$) values depends on both the gradient and the absolute value of the rotation velocity. Beyond R=8.5 kpc, Model A coincides with the $\sech^{2}$ model, hence the density values are the same and difference is 0\%.(Table 1).

Second, in Model B we consider the effect of R-z coupling, a non-flat rotation curve and planar random motions, hence we term it as the most general and the complete one. For this case, we solve eq. (8) from R=2 to 22 kpc, and compare the results with those from the $\sech^{2}$ model (Table 1) with $\Delta_B$ denoting the \% difference between these two models. We note that up to R=10 kpc, $\rho_{0}$ is lower than that of $\sech^{2}$ model, except at R=7 kpc where the density is higher as a result of the high magnitude of  negative gradient present in the rotation curve at that radius. We find that the magnitude of change in $\rho_{0}$ is higher in the inner radii and lower toward the solar neighborhood region.
In contrast, the resulting values of the mid-plane density are higher than those from the $\sech^{2}$ model at radii beyond R= 12 kpc. This change increases with radius and gets very prominent, more than $\sim 30$\%, in the outer disc region, beyond R=16 kpc. This may be explained as follows. We  expect that in the outer disc region, where the surface density is low and the vertical distribution becomes an extended one, as found in Sarkar \& Jog (2018), the contribution of the radial gradient of potential may not be negligible compared to the vertical gradient in the Poisson equation. Further, the effect of R-z coupling and the planar random motions 
can have substantial effect on the vertical density distribution since the self-gravity of the disc becomes low. Therefore, in the outer disc region it becomes necessary to consider the full Poisson equation with the full Jeans equations as done here.

For the intermediate radial range from 8.5-12 kpc, the values of change are small ( a few \%), and at R=12 kpc, the change is the lowest, almost zero. This is by sheer coincidence of the various input parameters which allow for the $\sech^{2}$ model to be valid in the intermediate radial range. Hence a simple $\sech^{2}$ model in the solar neighbourhood assumed so far in the literature has been well-justified. We note that this range showing small change may vary slightly depending on the actual values of the input parameters used. The changes shown in this range (see $\Delta B$ values in Table 1) do not show a clear pattern with radius. We note that it is the complex interplay among the various kinematical terms present in the vertical and radial terms in the Poisson equation (in eq. 8) that
sets the value of the mid-plane density and hence the difference with $\sech^{2}$ model results, including the sign of the difference. Beyond R=7 kpc, the rotation curve is taken to be flat so the difference then is purely due to the kinematical terms involving the random motions and the cross term.

We have shown the plots for $\rho(z)$ vs z for radii 6, 8.5, 18 \& 22 kpc in Fig.1 from Model B and the $\sech^{2}$ model. We note that we have provided the $\rho_{0}$ values with higher decimal accuracy to facilitate comparison among various models, though the observed data is not known to this accuracy.
Interestingly, we find that on inclusion of these kinematical terms, the resulting $\rho(z)$ profiles are found to fit to a function of type $\sech^{2/n}$ with n varying with radius (as was also found for the case of multi-component model shown in Sarkar \& Jog (2018)- which did not consider these kinematical effects), instead of the typical of $\sech^{2}$ function.

\subsection{Effect on the scaleheight (HWHM) of the stellar disc}

\begin{table}
\caption{HWHM using  the $\sech^{2}$ model, model with only a non-flat, observed rotation curve (Model A) and the complete, general model (Model B)}
\label{table:2}
\centering
  \begin{tabular}{l l l l l l}
\hline \hline
Radius & $\mathrm{(HWHM)_{\sech^{2}}}$ & $\mathrm{{(HWHM)}_{A}}$ & $\Delta_{A}$ & $\mathrm{(HWHM)_{B}}$ & $\Delta_{B}$  \\
(kpc)  & pc & pc & \% & pc & \% \\
\hline
2.0	&	266.3	&	251.8	&	-5.4	&	311.9	&	+17.1\\
3.0	&	290.2	&	301.8	&	+4.0	&	377.5	&	+30.0\\
4.0	&	313.6	&	340.0	&	+8.4	&	403.3	&	+28.6 \\
5.0	&	341.9	&	362.1	&	+5.9	&	408.8	&	+19.6\\
6.0	&	370.8	&	366.4	&	-1.2	&	393.3	&	+6.1 \\
7.0	&	405.1	&	371.7	&	-8.2	&	389.0	&	-4.0 \\
8.5	&	456.7	&	--	&	0.0	&	467.4	&	+2.3 \\
10.0	&	515.4	&	--	&	0.0	&	527.3	&	+2.3\\
12.0	&	606.2	&	--	&	0.0	&	599.4	&	-1.1\\
14.0	&	708.3	&	--	&	0.0	&	663.6	&	-6.3\\
16.0	&	822.2	&	--	&	0.0	&	760.5	&	-7.5\\
18.0	&	1568.9	&	--	&	0.0	&	1260.7	&	-19.6\\
20.0	&	2721.8	&	--	&	0.0	&	1858.9	&	-31.7\\
22.0	&	4279.9	&	--	&	0.0	&	2515.3  &	-41.2\\
\hline	
\end{tabular}
\end{table}

We measure the disc scaleheight in terms of the half-width at half-maximum (HWHM) of the vertical density distribution and list the values in Table 2. 
The symbols $\Delta_A$ and $\Delta_B$ have the same meaning as in Table 1.
In Model A, when the gradient in the rotation curve is positive it makes the disc puff up, and when the gradient is negative it decreases the disc thickness, compared to that of $\sech^{2}$ model due to higher mid-plane density (as similar to the case for UGC 7321(Sarkar \& Jog, 2019)). Beyond R= 8.5 kpc, Model A coincides with the $\sech^{2}$ model, hence the difference is 0\%(Table 2).
\par
In Model B, the change in HWHM from that of the $\sech^{2}$ model is positive upto R=10 kpc, except at R=7 kpc, due to the high negative gradient present in the rotation curve, as discussed earlier in 3.1. We get negative changes beyond R=12 kpc and also note that the amount of change is again the least at R=12 kpc. The change in scaleheight values become as high as 30-40\% in the outer disc region. We note that the corresponding changes in $\rho_{0}$ and HWHM are compatible with each other in both the models. 
  
Our results show that as in the $\sech^{2}$ case, in the complete model (Model B) also, the vertical disc thickness increases with radii and gives rise to flaring in the outer Galaxy. This again shows (as in Sarkar \& Jog 2018, Sarkar \& Jog 2019 where the kinematical effects were not included) that flaring in the outer disc region is a generic phenomenon. We note that the absolute magnitude of flaring from R=2 to 22 kpc decreases by a factor of two (from 16 to 8) compared to the $\sech^{2}$ model. This shows the importance of considering various effects (as in eq.(8)) while studying the vertical structure.

\subsection{Addition of dark matter halo gravity in the outer Galaxy}

For a realistic multi-component, gravitationally coupled galactic disc (but without the kinematical effects included here), the vertical distribution of stars is constrained by gas and dark matter halo gravity in the inner and the outer disc respectively (Sarkar \& Jog, 2018). Here we add the halo gravity to Model B at R=18, 22 kpc, considering the disc to be effectively a gravitationally coupled stars plus halo system and solve for $\rho(z)$. The $\rho_{0}$ values increase by a factor of 2.3 at R=18 kpc and a factor of 3 at R=22 kpc, compared to the stars-alone disc. The HWHM values also get lowered correspondingly, by 
a factor of 2.2 and 3.1 respectively and this tends to decrease the flaring. 
We do not add the gas gravity in the eq.(8) to solve it in the inner Galaxy, as the modified equation will contain second order variation in gas surface density, which is not known to us from observations. 

\medskip
\section{Conclusions}
\noindent  We have studied theoretically the vertical structure of a galactic  stellar disc using the most general and complete model (Model B here) which considers the complete axisymmetric Poisson equation containing both the radial and vertical terms. These are calculated using the complete Jeans equations which contain a non-flat rotation curve, planar random motions, and a tilted velocity ellipsoid. Thus, in this treatment the R and z motions are taken to be coupled. We found that the mid-plane density ($\rho_{0}$) is lower than that from the typical $\sech^{2}$ model at the inner radii and becomes higher from R=12 kpc onward. The HWHM values also change accordingly,i.e., higher at the inner radii and lower from R=12 kpc. Interestingly, by sheer chance, for the observed input parameters in the small radial range in the solar neighborhood $\sim 7-12$ kpc, the changes are very small ($<$ a few $\%$) whereas in outer disc region they are very prominent ($\sim 40\%$). This is due to the low density and hence low self-gravity and extended vertical disc distribution in the outer Galaxy.

\noindent Thus our work shows that the standard isothermal model resulting in a $\sech^{2}$ vertical density distribution (Spitzer 1942) 
for a thin, self-gravitating galactic disc turns out to be well-justified in the solar neighbourhood, but not in the outer disc.
Hence, one must include all the terms as discussed above to get the correct vertical distribution in the following three general cases of low density region, namely in the outer Galaxy, at high z and for a thick disc. The predicted changes due to the various kinematic effects studied here can now be verified with the new, accurate data for example from GAIA or LAMOST.
\medskip

\noindent Acknowledgements:
 We thank the anonymous referee for constructive and helpful comments. S.S. thanks CSIR for a fellowship, and C.J. thanks the DST 
 for support via J.C. Bose fellowship (SB/S2/JCB-31/2014).

\bigskip

\noindent {\bf {References}}
\medskip

\noindent Bahcall J., 1984, ApJ, 276, 156

\noindent Banerjee, A., Jog, C.J., Brinks, E., Bagetakos, I., 2011, MNRAS, 415, 687

\noindent Binney J., Tremaine, S., 1987, Galactic Dynamics ( Princeton: Princeton Univ. Press)

\noindent Bovy J.,  Tremaine, S., 2012, ApJ, 756, 89

\noindent Cappellari M., 2008, MNRAS, 390, 71

\noindent Cappellari M., 2019, submitted to MNRAS (arXiv:1907.09894)

\noindent Dehnen W., Binney, J., 1998, MNRAS, 298, 387

\noindent Dickey J. M. 1996, in Unsolved Problems of the Milky Way, eds. L. Blitz, \&
P. Teuben (Dordrecht: Kluwer), IAU Symp., 169, 489

\noindent Eilers A-C., Hogg D.W., Rix H.W., Ness M.K, 2019, ApJ, 871, 120
 
\noindent Everall A., Evans N.W., Belokurov V., Schönrich R., 2019, MNRAS, 489, 910

\noindent Hagen J.H.J., Helmi A., 2018, A\&A, 615, A99

\noindent Hagen J.H.J., Helmi A., de Zeeuw P.T., Posti L., 2019, A\&A,629, A70

\noindent Kamphuis J. J., 1993, PhD Thesis, University of Groningen

\noindent Lewis J. R., Freeman K. C. 1989, AJ, 97, 139

\noindent Mera D., Chabrier, G., Schaeffer, R., 1998, A\&A, 330, 953

\noindent Mignard F., 2000, A\&A, 354, 522

\noindent Mihalas D., Routly P.M., 1968, Galactic Astronomy

\noindent Narayan, C.A., Jog, C.J. 2002, A\&A, 394, 89

\noindent Narayan, C.A., Saha, K., Jog, C.J. 2005, A\&A, 440, 523

\noindent Nitschai M.S., Cappellari M., Neumayer N., 2019, MNRAS

\noindent Oort, J.H.  1932, BAN, 6, 349

\noindent Sánchez-Salcedo F.J., Flynn C., de Diego J.A., 2016, ApJ, 817, 13

\noindent Sarkar S., Jog C.J., 2018, A\&A, 617, A142

\noindent Sarkar S., Jog C.J., 2019, A\&A, 628, A58

\noindent Sofue Y., 2012, PASJ, 64, 75

\noindent Spitzer L., 1942, ApJ, 95, 329

\noindent Wegg C., Gerhard O., Bieth,M., 2019, MNRAS, 485, 3296

\end{document}